\newcommand{\mass}[1]{\ensuremath{#1\, \mathrm{M}_\odot}}

\newcommand{\chem}[2]{\ensuremath{^{#2}\kern-0.8pt\mathrm{#1}}}
\newcommand{\reac}[6]{\ensuremath{\,^{#2}\kern-0.8pt\mathrm{#1}\,({#3}\,,{#4})\,{}^{#6}\kern-0.8pt\mathrm{#5}\,}}

\newcommand{\PL}{PL}

\documentclass[letter]{aa} 

\usepackage{graphicx}
\usepackage{enumerate}
\usepackage{txfonts}
\usepackage{natbib}
\usepackage{ulem}
\usepackage{color}

\begin{document} 

   \title{Stellar variability in open clusters}

   \subtitle{II. Discovery of a new period-luminosity relation in a class of fast-rotating pulsating stars in NGC~3766\footnote{Based on observations made with the FLAMES instruments on the VLT/UT2 telescope at the Paranal Observatory, Chile, under the program ID 69.A-0123(A)}}

   \author{N. Mowlavi\inst{1},
           S. Saesen\inst{1},
           T. Semaan\inst{1},
           P. Eggenberger\inst{1},
           F. Barblan\inst{1},
           L. Eyer\inst{1},
           S. Ekstr\"om\inst{1},
          \and
           C. Georgy\inst{1}
          }

   \institute{Institute of Astronomy, University of Geneva, 51 chemin des Maillettes, 1290 Versoix, Switzerland\\
              \email{Nami.Mowlavi@unige.ch}
             }

   \date{Received 23 June 2016; accepted 27 September 2016}

 
  \abstract
   {
   Pulsating stars are windows to the physics of stars enabling us to see glimpses of their interior.
   Not all stars pulsate, however.
   On the main sequence, pulsating stars form an almost continuous sequence in brightness, except for a magnitude range between $\delta$ Scuti and slowly pulsating B stars.
   Against all expectations, 36 periodic variables were discovered in 2013 in this luminosity range in the open cluster NGC~3766, the origins of which was a mystery.
   }
   {
    We investigate the properties of those new variability class candidates in relation to their stellar rotation rates and stellar multiplicity.
   }
   {
    We took multi-epoch spectra over three consecutive nights using ESO's Very Large Telescope.
   }
   {
    We find that the majority of the new variability class candidates are fast-rotating pulsators that obey a new period-luminosity relation.
    We argue that the new relation discovered here has a different physical origin to the period-luminosity relations observed for Cepheids.
    }
   {
    We anticipate that our discovery will boost the relatively new field of stellar pulsation in fast-rotating stars, will open new doors for asteroseismology, and will potentially offer a new tool to estimate stellar ages or cosmic distances.
   }


\titlerunning{New period-luminosity relation in a class of fast-rotating pulsating stars}
\authorrunning{N. Mowlavi et al.}
\maketitle


\section{Introduction}
\label{Sect:introduction}

The existence of a relation between intrinsic quantities in a star is a goldmine in which to probe the physics of stars, enabling us to derive key information on the stars that would otherwise be difficult to obtain.
Some pulsating stars, in particular, are known to obey a relation between the period $P$ of their photometric variability and their luminosity $L$.
This period-luminosity (\PL) relation essentially links the characteristic dynamical timescale of a star to its mean density.
This was first discovered at the beginning of last century for Cepheids in the Small Magellanic Cloud \citep{LeavittPickering12}, a discovery that eventually led to the provision of a ladder with which to measure cosmic distances \citep{Hubble25,Fernie69}.
In the late sixties, the relation enabled  the masses of Cepheids to be estimated using stellar pulsation models.
Surprisingly, the masses derived in this way were found to be smaller by a factor of up to two
than the masses determined by stellar evolution models \citep{Stobie69,Rodgers70}.
The discrepancy kept astrophysicists active for decades \citep{BonoCaputoCastellani06}, with a satisfactory solution  only recently being proposed from the study of the impact of stellar rotation on the evolution of Cepheid progenitors \citep{AndersonEkstromGeorgy_etal14}.

The new \PL\ relation reported here is potentially of a new kind.
It is found in a new class of periodic variables that  burn hydrogen in their core (main-sequence stars), discovered in the young ($\sim$20 million years) open cluster NGC~3766, which is located at about 6500 light years from the Sun \citep[][hereafter Paper~I]{MowlaviBarblanSaesen_etal13}.
Different classes of pulsating stars have been identified  on the main sequence \citep{Christensen-Dalsgaard04,AertsChristensen-DalsgaardKurtz10,CatelanSmith15}.
They form an almost continuous sequence in increasing brightness, except for a magnitude range between $\delta$ Scuti and slowly pulsating B (SPB%
\footnote{The term SPB star denotes in this Letter the classical SPB stars defined by the (classical) limits of the SPB instability strip based on models of non-rotating stars.
}%
) stars \citep{Pamyatnykh99,Saio14}.
Against all expectations, the new periodic variables in NGC 3766 were found to be in this luminosity range.
They are more than $\sim$2.5 times more massive than the Sun, and more than $\sim$40 times brighter.
Their light fluxes are modulated at the per mille level on timescales of about 2 to 15 hours.
This is shorter than the periods of light modulation detected in the brighter SPB stars, and longer than those observed in the fainter $\delta$ Scuti stars.
No standard scenario could explain the properties of these new variables.
These properties are a challenge for standard stellar models \citep{MowlaviSaesenBarblan_etal14a}, but they could be due to the combined effect of rotation and pulsation (\citealt{SalmonMontalbanReese_etal14}; see also Sect.~7.2 of Paper~I).

Unexpected detections of periodic variables in the instability gap between the $\delta$~Scuti and SPB instability strips have already been reported in the literature \citep[see Sect.~7.3 of Paper~I for a discussion of Maia and anomalous "rapidly rotating SPB" stars, and][for the case of hot $\gamma$~Doradus stars]{BalonaEngelbrechtJoshi_etal16}.
The link between these anomalous variables, the origins of which are as yet unknown, and the new variables reported in NGC~3766, is not obvious.
All the new variables may not even  have a unique origin.
It was suggested in Paper~I that they may be divided in two subgroups, a separation confirmed by the present study.

To understand the nature of the new variables in NGC~3766, we conducted a spectroscopic observation program  on ESO's Very Large Telescope at the Paranal Observatory, Chile.
Multi-epoch spectra were taken for 183 stars, comprised  of all the new variability class and SPB candidates, as well as other stars in the magnitude range of interest.
In this letter, we analyze the spectroscopic properties of the new variable stars (Sect.~\ref{Sect:dataAnalysis}).
We show evidence for the discovery of a \PL\ relation of a potentially new kind that is obeyed by the fast-rotators among the new variables (Sect.~\ref{Sect:PLrelation}).
The origin of the new relation is investigated in Sect.~\ref{Sect:origin} and some astrophysical implications, together with conclusions, are presented in Sect.~\ref{Sect:conclusions}.
Appendices \ref{Appendix:photometry} to \ref{Appendix:MRLrelations} provide more details on the photometric data, on the identification of binary stars, on the computation of stellar rotation rates, and on the interpretation of the \PL\ relation.
A table with the data pertaining to the new variables obeying the \PL\ relation is provided in Appendix~\ref{Appendix:rotationRates}.

\section{Star selection and properties}
\label{Sect:dataAnalysis}

\begin{figure}
  \centering
  \includegraphics[width=1\columnwidth]{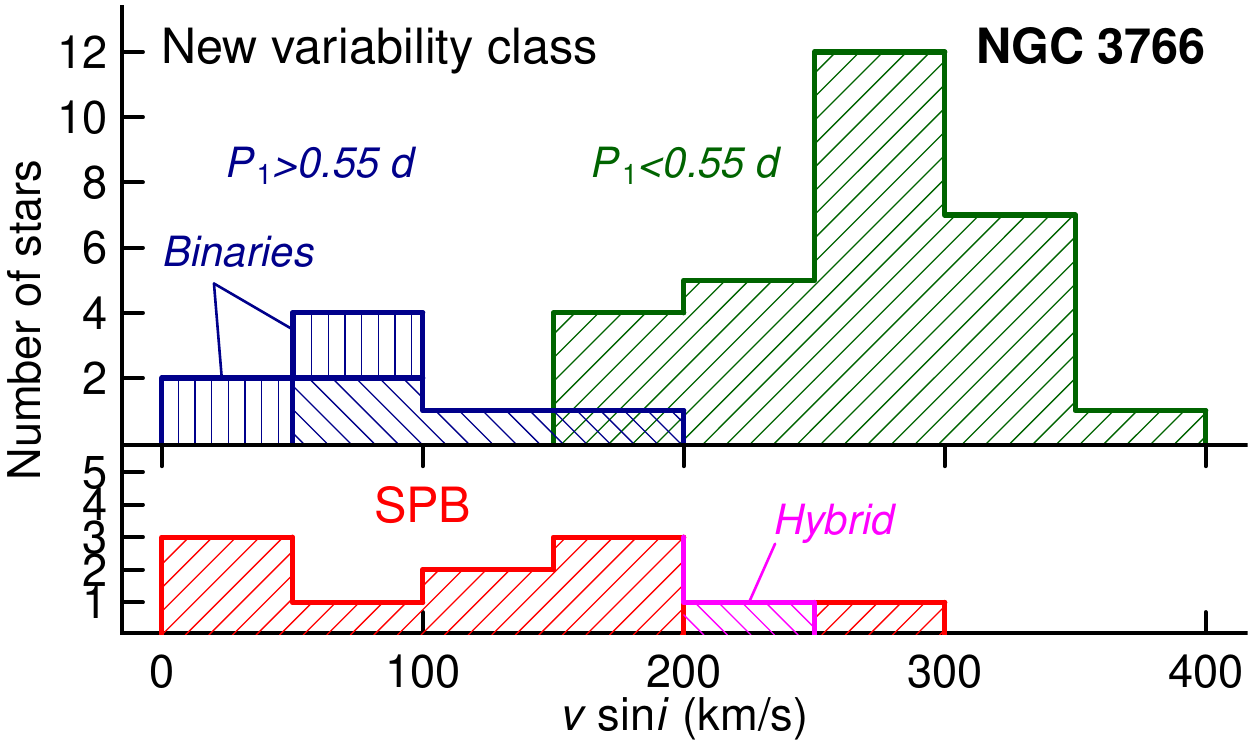}
  \caption{\textbf{Top panel:} Histogram of the projected rotational velocities of the new variability class candidates in NGC~3766, divided into candidates with their dominant photometric period $P_1>0.55$~d (135$^\circ$ and vertical shaded blue) and $P_1<0.55$~d (45$^\circ$ shaded green).
           Confirmed binaries are vertically shaded.
           \textbf{Bottom panel}: same as top panel, except for the SPB candidates (45$^\circ$ shaded red) and the hybrid star (135$^\circ$ shaded magenta).
           }
\label{Fig:histogramVsini}
\end{figure}

\begin{figure}
  \centering
  \includegraphics[width=1\columnwidth]{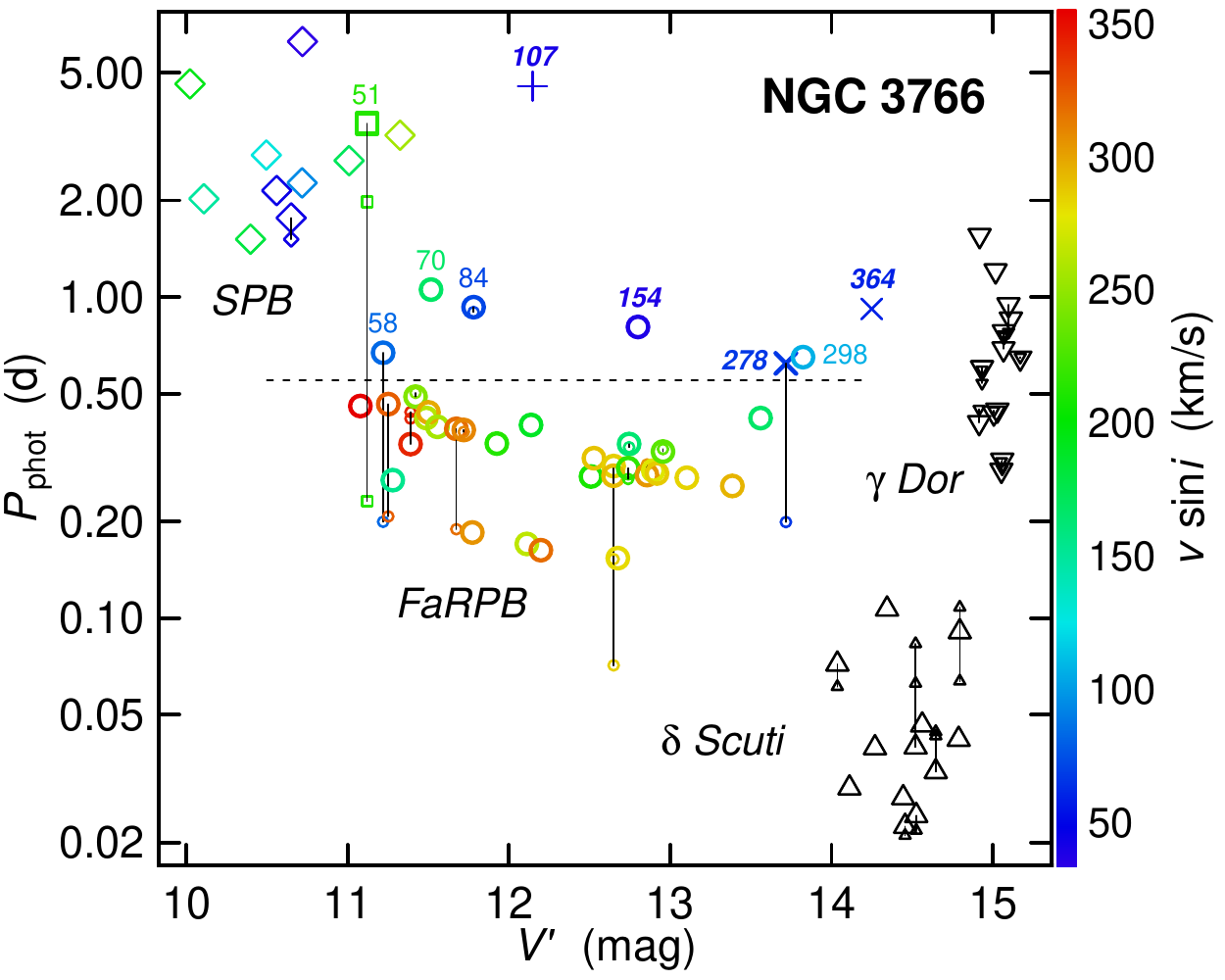}
  \includegraphics[width=1\columnwidth]{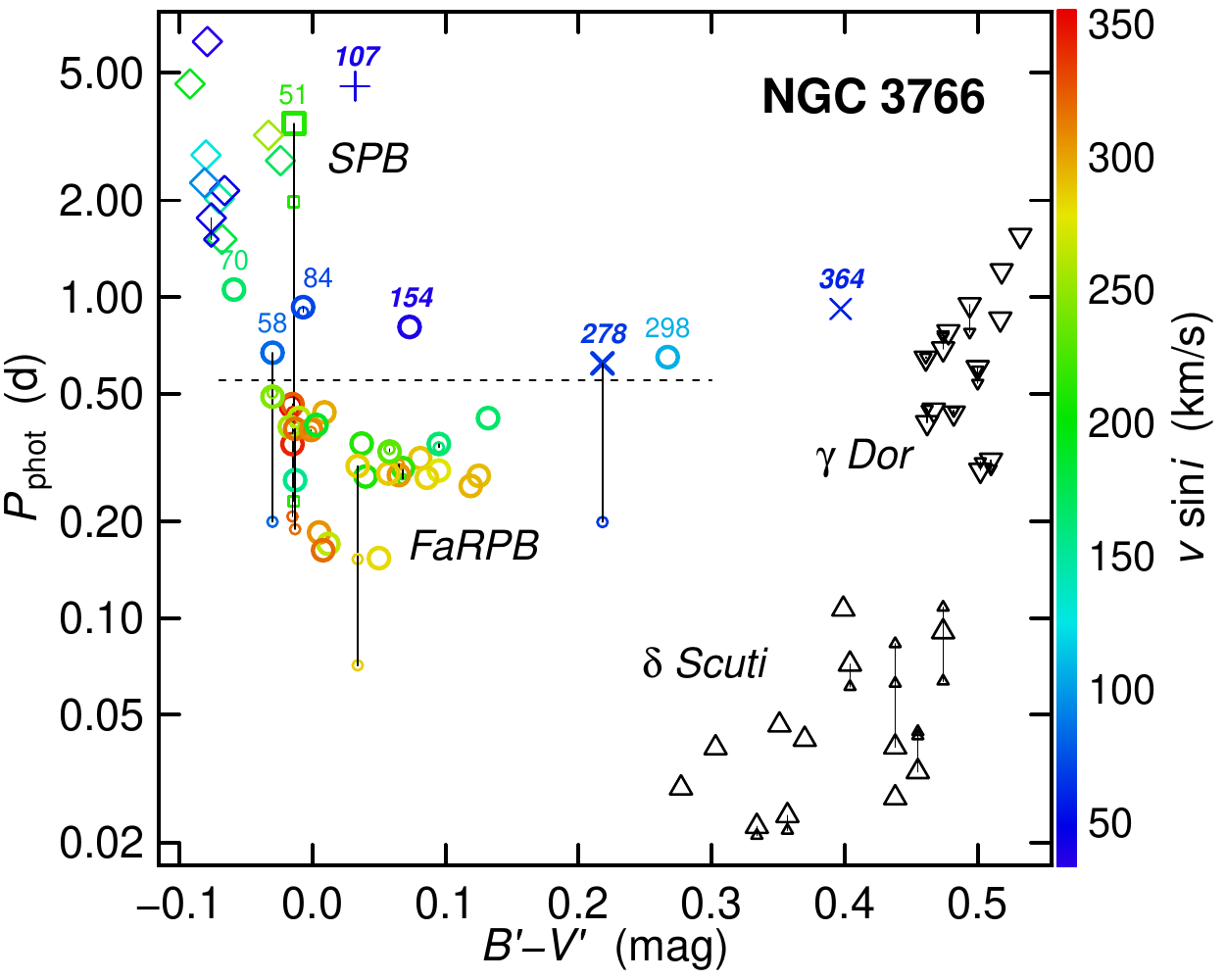}
  \caption{Period-magnitude (top) and period-color (bottom) diagrams of the periodic variable candidates of NGC~3766 brighter than 15.5~mag.
           The photometric periods $P_\mathrm{phot}$ and apparent magnitudes are taken from Paper~I.
           The projected stellar rotational velocities derived in this work are color-coded according to the color-scales on the right of the diagrams, or are in black when no spectrum is available.
           SPB candidates are marked with diamonds, the new variability class candidates (including FaRPB stars) with circles, $\delta$~Scuti stars with upward triangles, and $\gamma$~Dor stars with downward triangles.
           The unique hybrid SPB/FaRPB star is marked with squares.
           Photometric periods resulting from orbital motion in a binary system are plotted with plus sign (SB1-type binary) or cross  (SB2-type binary) signs.
           Solid vertical lines link periods within each multiperiodic star, with the second and third periods drawn with smaller indicators.
           A dashed line is added at $P_\mathrm{phot}=0.55$~d as a guide for the eye.
           New variability class candidates with periods greater than this limit are labeled with their star IDs next to the indicators, the four stars in binary systems being labeled in bold italic.
  }
\label{Fig:periodMagnitude}
\end{figure}

We consider all stars of the new variability class and SPB stars identified in Paper~I, except star 295\footnote{
The star IDs adopted here are the ones used in Paper~I.
},
which is strongly reddened and possibly does not belong to NGC~3766 (see Appendix~\ref{Appendix:photometry}).
Their photometric properties, magnitudes, and periods are likewise taken from Paper~I.
The period distribution (Fig.~19 in Paper~I) showed a bulk of periods between 0.1~d and $\sim$0.5~d for
the new variability class candidates, with a few outliers having periods that were greater than 0.7~d.

Multi-epoch spectra were taken at the VLT UT2-Kueyen during three consecutive nights starting on April 17, 2014.
A total of 3 437 spectra (a mean of 19 spectra per target) were taken in the [3964, 4567]~{\AA} wavelength range with the FLAMES/GIRAFFE instrument (LR2 mode, spectral resolution of R$\sim$6000), and one additional spectrum was taken the last night for each of the 183 stars in the [6438, 7184]~{\AA} wavelength range (LR6 mode, R$\sim$8500) to identify Be stars from the presence of emission in their H$_\alpha$ line.
The sample of stars observed with FLAMES/GIRAFFE comprises the 36 new variability class candidates, the 12 SPB candidates, and the one hybrid candidate (that shares properties of both the new variability class and SPB stars), which were identified in 2013, as well as a random selection of stars in the magnitude range of interest that did not show periodic photometric variability.
The spectroscopic properties of the full sample of observed stars will be presented in Semaan et al. (in prep.); here we focus on the new variables and SPB stars.

Our goal is twofold with the spectra.
First we  want to identify the binary stars among the new variable class and SPB candidates, to check whether the periodic variability could originate from the orbital motion of tidally distorted binary components.
We then want to characterize the stellar rotational properties of all these variable stars.
The details of the analysis are given in Appendices \ref{Appendix:binaries} and \ref{Appendix:rotationRates}.
In summary, four of the new variability class candidates are found to be in binary systems.
Their photometric periods are all larger than 0.55 d (see Figs~\ref{Fig:histogramVsini} and \ref{Fig:periodMagnitude}, which are further analyzed in Sect.~\ref{Sect:PLrelation}).
This result confirms the suspicion, which was raised in 2013, that the origin of at least some of the large photometric periods among the new variability class candidates may be different to the origin of the newly discovered variability class.
All new variability class candidates with photometric periods smaller than 0.55~d, on the other hand, are found to be fast rotators (see Fig.~\ref{Fig:periodVSini} in the Appendix).
We analyze these in the next sections.

\section{The new period-luminosity relation}
\label{Sect:PLrelation}

The analysis of the spectra reveals that the majority of the new variability class candidates are fast-rotators (top histogram in Fig.~\ref{Fig:histogramVsini}).
As  discussed in Sect.~\ref{Sect:origin}, it is suggested that pulsation  plays a role in their photometric variability, in addition to fast rotation, and we hereafter call them fast-rotating pulsating B (FaRPB) stars.
They are distinct from SPB stars due to their smaller periods, which are below 0.55~d (Fig.~\ref{Fig:periodMagnitude}),  their faster rotation rates (Fig.~\ref{Fig:histogramVsini}),  their lower luminosities (Fig.~\ref{Fig:periodMagnitude}), and their generally smaller photometric variability amplitudes (see X-axis distribution in Fig.~\ref{Fig:RvVarAmplitude}).
The fact that so many FaRPB stars have projected rotation velocities $v \sin(i) > 200$ km/s ($v$ is the equatorial speed and $i$ the inclination angle of the rotation axis with respect to the line of sight) strongly suggests that they are seen close to equator-on.

\begin{figure}
  \centering
  \includegraphics[width=1\columnwidth]{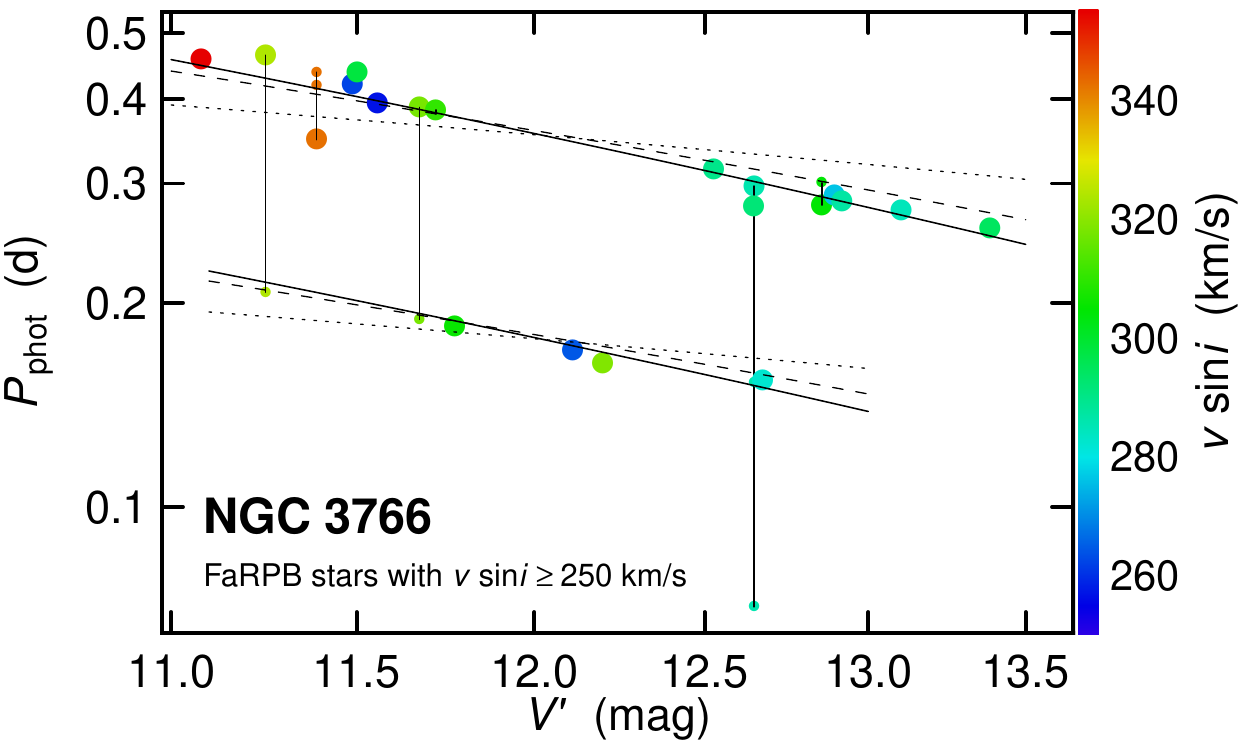}
  \caption{Period-magnitude diagram for FaRPB stars with $v\sin i \geqslant 250$km/s.
           Solid vertical lines link periods within each multi-periodic star, with the second and third periods drawn with smaller indicators.
           The two solid lines locate the \PL\ relation discovered in this paper from a combined fit to the data on the two sequences.
           The dashed and dotted lines illustrate relations that would result from analytical arguments linking the \PL\ relation to stellar rotation and Cepheid-like pulsation, respectively (see text).
           The $v\sin i$ values are color-coded according to the color scale on the right of the figure.
           }
\label{Fig:PLrelation}
\end{figure}

Remarkably, FaRPB stars are seen to obey two sequences in the period-magnitude diagram (Fig.~\ref{Fig:PLrelation}).
FaRPB stars rotating faster than $v_\mathrm{min,FaRPB}=250$~km/s obey, to first order, linear relations between the logarithm of periods and $V$ magnitudes.
The periods on the first (upper, with larger periods) sequence are about twice as long as those on the second (lower, with smaller periods) sequence at any given magnitude.
Assuming a factor of two between the periods on the two sequences, a linear fit to the combined data of both sequences leads to a relation between the photometric period $P_\mathrm{phot}$ and apparent magnitude%
\footnote{The uncalibrated $V'$ magnitudes displayed in the figures in this Letter are shown in Appendix~\ref{Appendix:photometry} as being equivalent to Johnson magnitudes.
}
$V$ given by, for the first sequence,
\begin{equation}
  \log(P_\mathrm{phot}) = (-0.11 \pm 0.01) \; (V-12) - (0.449 \pm 0.006),
\label{Eq:PLrelation}
\end{equation}
with $P_\mathrm{phot}$ expressed in days.
The relation is shown in solid lines in Fig.~\ref{Fig:PLrelation}.
The value of the slope is not sensitive to the adopted lower limit $v_\mathrm{min,FaRPB}$, as long as $v_\mathrm{min,FaRPB} > 170$~km/s.
Below this limit, two of three stars that have $v\sin i$ between 150~km/s and 170~km/s are located far out of the sequences (see Appendix~\ref{Appendix:rotationRates}).

Four additional properties of the sequences emerge from the data shown in Fig.~\ref{Fig:PLrelation}:
(i) There is a small dispersion within each sequence.
The fact that some multi-periodic FaRPB stars have several periods that fall in the same sequence indicates that the dispersion is intrinsic to the mechanism at the origin of the periodic variability.
(ii) There are multi-periodic stars that have periods in both sequences.
(iii) In the three multi-periodic FaRPB stars that have periods in both sequences, the dominant period, i.e. with the largest amplitude of photometric variability, is always on the first sequence.
(iv) Finally, the second sequence is much less populated than the first one (there are about two times less stars on the second sequence than on the first sequence if we consider FaRPB stars with $v\sin i > 250$~km/s, and about three times less when taking all FaRPB stars with $v\sin i>150$~km/s), and does not extend to as faint stars as the first sequence does.

\section{Origin of the relation}
\label{Sect:origin}

Two processes can be envisaged to explain the existence of the \PL\ relation discovered in Sect.~\ref{Sect:PLrelation}.
The first is the mechanism at play in Cepheid-like pulsating stars, for which the pulsation period $P_\mathrm{puls}$ is related to the stellar mean density $\bar{\rho}$ through $P_\mathrm{puls} \propto 1/\!\!\sqrt{\bar{\rho}}$.
The second process is stellar rotation, with a direct relation between photometric period and rotation period $P_\mathrm{rot} = 2\pi R_\mathrm{eq}/v$, $R_\mathrm{eq}$ being the equatorial radius.

Qualitative arguments using $L \propto M^\alpha$ and $R \propto M^\beta$ dependencies between stellar luminosity $L$, radius $R$, and  mass $M$ of main sequence stars favor a link with stellar rotation.
Taking $\alpha\!\simeq\!3.8$, $\beta\!\simeq\!0.5$, and first order approximation $V \simeq 0.6 M_\mathrm{bol} + \mathrm{cte}$ to convert bolometric magnitude $M_\mathrm{bol}$ to $V$ magnitude (see Appendix~\ref{Appendix:MRLrelations}), relations of the form
\begin{eqnarray}
  \log(P_\mathrm{puls}) & \simeq & - \frac{3\beta-1}{3\alpha} \, V + \mathrm{cte} \;\; \simeq \;\; -0.044 \, V  + \mathrm{cte}
\label{Eq:Ppuls}\\
  \log(P_\mathrm{rot})  & \simeq & - \frac{2}{3} \frac{\beta}{\alpha} \, V + \mathrm{cte} \;\; \simeq \;\; -0.088 \, V  + \mathrm{cte}
\label{Eq:Prot}
\end{eqnarray}
are expected (for our investigations, we assume in Eq.~\ref{Eq:Prot} that $v$ has about the same value in all our FaRPB stars, $v\sin i$ being equal to $300\pm 50$~km/s for more than 65\% of the FaRPB stars, see Fig.~\ref{Fig:histogramVsini}).
Despite the qualitative nature of the reasoning, comparison of the slopes in Eqs.~\ref{Eq:Ppuls} and \ref{Eq:Prot} with the observed slope given in Eq.~\ref{Eq:PLrelation} suggests a link of the \PL\ relation with stellar rotation, rather than with a Cepheid-like pulsation mechanism (linear fits to the combined data using these slopes are shown in Fig.~\ref{Fig:PLrelation}).
In this scenario, a small non-linearity of the relation is expected owing to the non-linear dependence of $V$ on $M_\mathrm{bol}$ and the non-identical values of $v$ in FaRPB stars.

Another argument in favor of a rotation origin of the relation comes from the 1/2 ratio between the periods of the second and first sequences.
This value is not compatible with typically greater than 1/2 period ratios of double mode Cepheid-like pulsators \citep{Moskalik14}, which
strongly supports a stellar rotation origin of the relation.

\vskip 2mm
The above analysis of the \PL\ relation suggests a direct link between the photometric periodicity and the stellar rotation rate.
This can result from the presence of luminosity/temperature structures at the surface of rotating stars.
Stellar spots sustained by magnetic fields would be the first  scenario to come to mind for such co-rotating structures.
The properties of the multi-periodic FaRPB stars, however, do not seem to support this type of scenario.
The three periods of multi-periodic star 62 at $V$ = 11.39 mag, for example, with periods of 0.348156~d, 0.41805~d, and 0.43796~d, would imply an improbable surface differential rotation of 25\% if they were to be explained by the presence of three spots, each located at different latitudes at the surface of the star.

Pulsation in rotating stars can also give rise to observed frequencies close to the rotation frequency, if the pulsation frequencies $\omega_\mathrm{corot}$ in the co-rotating frame are much smaller than the rotation frequency $\Omega$.
An observer in the inertial frame would then measure pulsation frequencies $\omega_\mathrm{obs} = \omega_\mathrm{corot} - m \Omega$ ($m$ being the azimutal order) close to an integer multiple of the rotation frequency.
Frequencies $\omega_\mathrm{corot} \ll \Omega$ are actually predicted by stellar models of pulsation in fast-rotating stars for high-order prograde sectoral modes \citep{Townsend05a}, i.e. for $m = -\ell$ where $\ell$ is the angular degree of the mode.
In this scenario, frequencies detected in the upper sequence would belong to $(\ell, m) = (1, -1)$ modes and those in the lower sequence to $(\ell, m) = (2, -2)$ modes.
These g-mode pulsations are predicted as being confined to the equator in fast-rotating stars \citep{UshomirskyBildsten98}, a prediction that is in agreement with the absence of small projected rotational velocities measured in FaRPB stars in NGC 3766.

Pulsation in fast rotating stars is thus the most likely explanation of the periods detected in FaRPB stars, with observed periods close to an integer fraction of the rotation period.
We would then expect to observe more sequences  if higher azimuthal orders are excited.
It is interesting to note that the very small period of 0.0712331~d detected in triple-periodic star 142 (its two other photometric periods are 0.29739 d and 0.152541 d) could possibly be the unique example in our photometric data of a third sequence corresponding to $m = -4$.
Quantitative agreement between predictions and observations, however, awaits further studies from both model predictions \citep{SalmonMontalbanReese_etal14} and additional observations.

\vskip 2mm
While the above analysis points to a link between FaRPB photometric periods and rotation rates, the very origin of the new \PL\ relation remains a mystery.
In an open cluster, all stars have the same age and the magnitude of a star traces its mass or radius.
The existence of a \PL\ relation in FaRPB stars thus suggests a relation between period and stellar radius.
This would imply a dependency of the \PL\ relation to the age of the cluster since stellar radius changes with age.
Observation of FaRPB stars in young stellar clusters other than NGC~3766 could provide key information in this respect.

\section{Conclusions}
\label{Sect:conclusions}

The discovery of the FaRPB \PL\ relation opens doors to the potential provision of a new tool to estimate stellar ages or distances if the universality of the age-dependent relation is confirmed.
To achieve this, FaRPB stars must be identified in additional young clusters of various ages, and their \PL\ relations studied and calibrated with the cluster's age.
If this can be achieved, the calibrated age-dependent relation can serve to either determine the age of a FaRPB star, if its distance to the Sun is known, or to determine the distance of a FaRPB star, if its age is known.
The former option is particularly appealing in view of ESA's Gaia space mission, currently in operation, which aims to provide distances for more than one billion stars in our Galaxy \citep{deBruijne12}.
The latter option, on the other hand, can be applied to stars belonging to, for example, stellar clusters, the ages of which can be determined using alternative techniques.

On the theoretical side, a more thorough investigation of pulsation frequencies predicted by models of very fast-rotating stars may open doors to asteroseismology studies on these stars, eventually providing a better insight on the origin of the new \PL\ relation.
We thus anticipate our work boosting the young field of stellar pulsation in fast-rotating stars.

Finally, we must mention the similarity of some properties of our FaRPB stars with properties of some Be stars, a class of actively studied fast-rotating B-type stars that show emission lines in their spectra \citep{RiviniusCarciofiMartayan13}.
Several Be stars have been observed from space-based MOST and CoRoT telescopes and found to contain a rich frequency spectrum that is distributed in distinct groups around so-called mean periods that are a factor of two of one another \citep{WalkerKuschnigMatthews_etal05,Gutierrez-SotoFloquetSamadi_etal09,Saio13}.
This feature is reminiscent of the properties of the two sequences discovered in our FaRPB stars in NGC 3766.
Actually, star 167, which is among the faintest and coolest of the FaRPB stars, was found to be a Be star owing to the presence of emission lines in its VLT spectrum.
If a link between the two classes of stars can be confirmed, the study of the properties of FaRPB stars will enable  a better understanding of Be stars, and vice-versa.

\begin{acknowledgements}

We thank the anonymous referee for constructive comments that helped improve the article.
\end{acknowledgements}

\bibliographystyle{aa}
\bibliography{bibTex}

\begin{appendix}

\section{Photometric properties}
\label{Appendix:photometry}

\begin{figure}
  \centering
  \includegraphics[width=1\columnwidth]{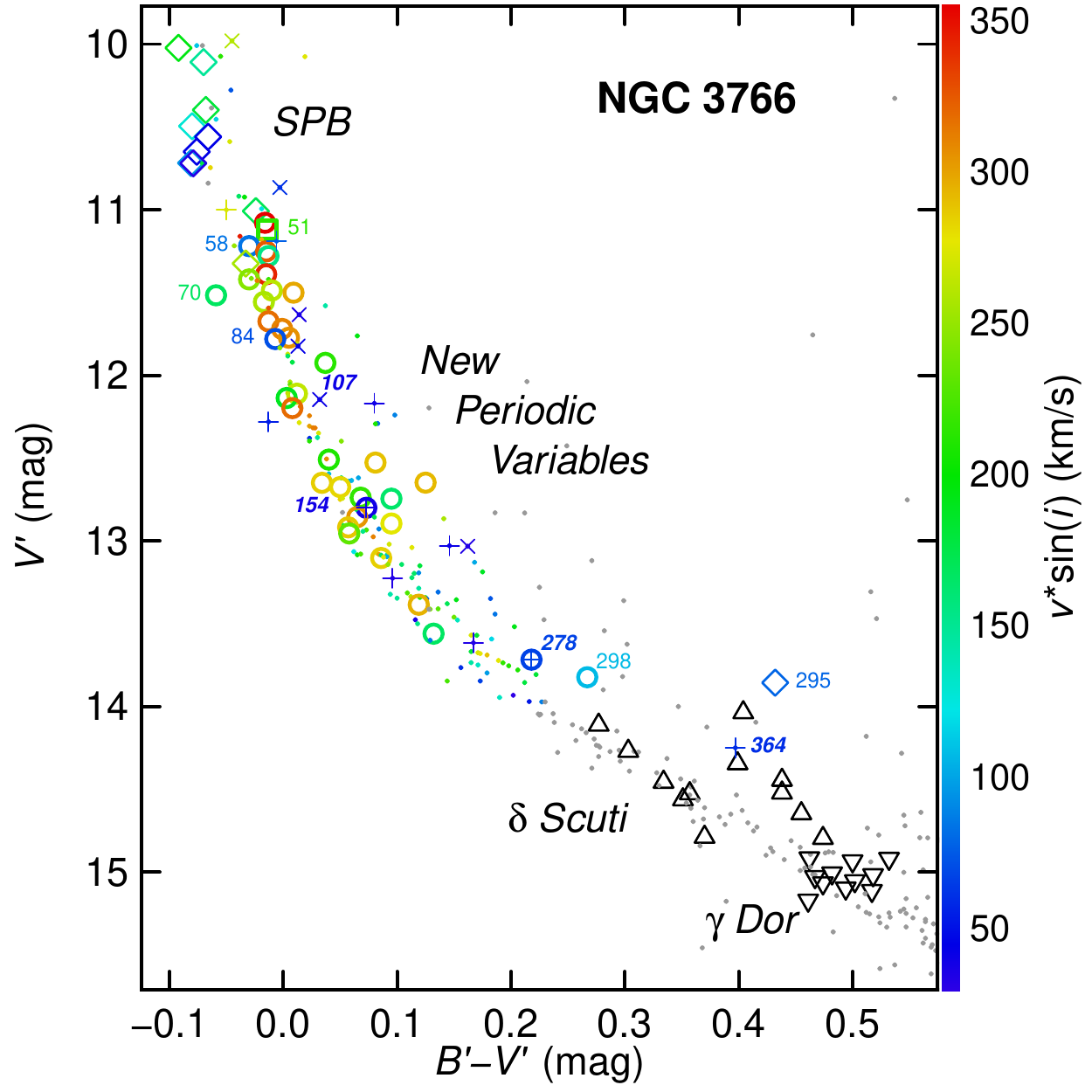}
  \includegraphics[width=1\columnwidth]{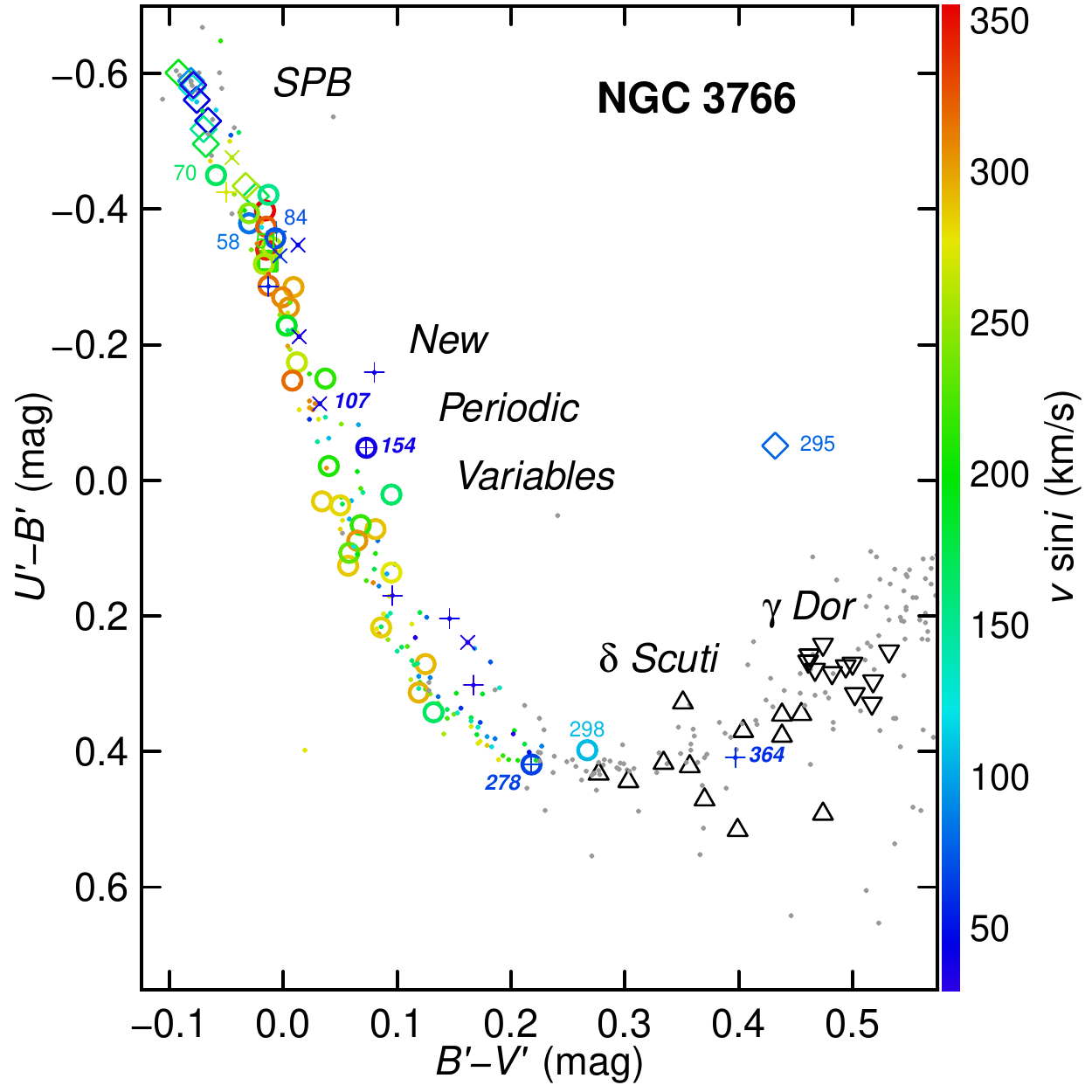}
  \caption{Color-magnitude (top) and color-color (bottom) diagrams of NGC~3766, limited to magnitude and color ranges of interest to the periodic stars studied in this paper.
           Indicator symbols, colors and labels have the same meanings as in Fig.~\ref{Fig:periodMagnitude}.
           Stars for which no spectra is available in this program are plotted in black for periodic variables and in gray for non-periodic stars. Stars labeled in Fig.~\ref{Fig:periodMagnitude} are reported in the diagrams, as well as the highly reddened non-cluster member star 295.
  }
\label{Fig:CM-CC}
\end{figure}

\begin{figure}
  \centering
  \includegraphics[width=0.85\columnwidth]{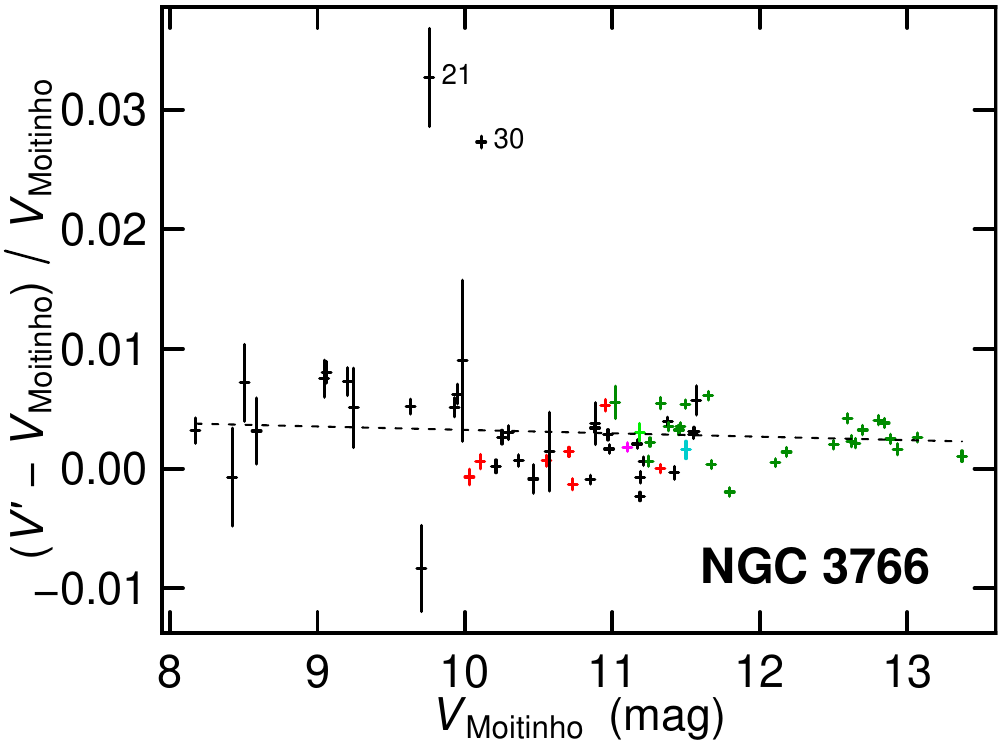}
  \caption{$V'$ magnitudes from Paper~I compared to Johnson $V$ band magnitudes from \cite{MoitinhoAlfaroYun_etal97}, the latter labeled $V_\mathrm{Moitinho}$ in the figure.
           Red indicators identify SPB candidates, and green indicators, new variability class candidates.
           The vertical lengths of the indicators equal the uncertainties in $V'$, taken to be the largest value between the $V'$ error bar from Paper~I and the standard deviation of $V'$ light curve.
           The horizontal lengths equal the uncertainties in $V_\mathrm{Moitinho}$ taken from Table~3 of \cite{MoitinhoAlfaroYun_etal97}.
           A weighted fit to the data, excluding stars 21 and 30 because they show bursts in their light curves, leads to $V' = (1.000 \pm 0.004) V_\mathrm{Moitinho} + (0.031 \pm 0.041)$~mag.
           The fit, converted to the relative deviation of $V'$ with respect to $V_\mathrm{Moitinho}$, is shown with the dashed line in the figure.
  }
\label{Fig:MoitinhoV}
\end{figure}

All photometric data are taken from Paper I, where we reported the discovery of the new variability class.
The color-magnitude and color-color diagrams are shown in Fig.~\ref{Fig:CM-CC}, where the projected rotational velocities that were computed from the VLT spectra are reported in color.
One of the new variability class candidates, star 295, is seen in the color-color diagram to be highly reddened.
We are tempted to think that it does  not belong to the cluster, yet its radial velocity is compatible with that of the cluster.
Since its $V'$ magnitude is anyway affected by the reddening, we do not consider it in this study, and do not show it in the figures of this paper, except in Fig.~\ref{Fig:CM-CC}.

The non-calibrated $V'$ magnitudes reported here, taken from paper~I, are equivalent to Johnson $V$ magnitudes for the B- and A-type stars of interest in this work.
This is shown in Fig.~\ref{Fig:MoitinhoV}, which compares the uncalibrated $V'$ magnitudes of Paper~I with the Johnson $V$ band magnitudes determined by \cite{MoitinhoAlfaroYun_etal97}.
The comparison shows a systematic offset of less than 0.4\%.
The  $B'-V'$ and $U'-B'$ colors, on the other hand, are not equivalent to the Johnson $B-V$ and $U-B$ colors, respectively.
We do not use the colors published by \cite{MoitinhoAlfaroYun_etal97} because they do not take into account the intrinsic variability of the stars.
Rather, we keep the uncalibrated color values published in Paper~I, knowing that this does not impact on any of the conclusions in this paper.

\section{Binary stars identification}
\label{Appendix:binaries}

\begin{figure}
  \centering
  \includegraphics[width=1\columnwidth]{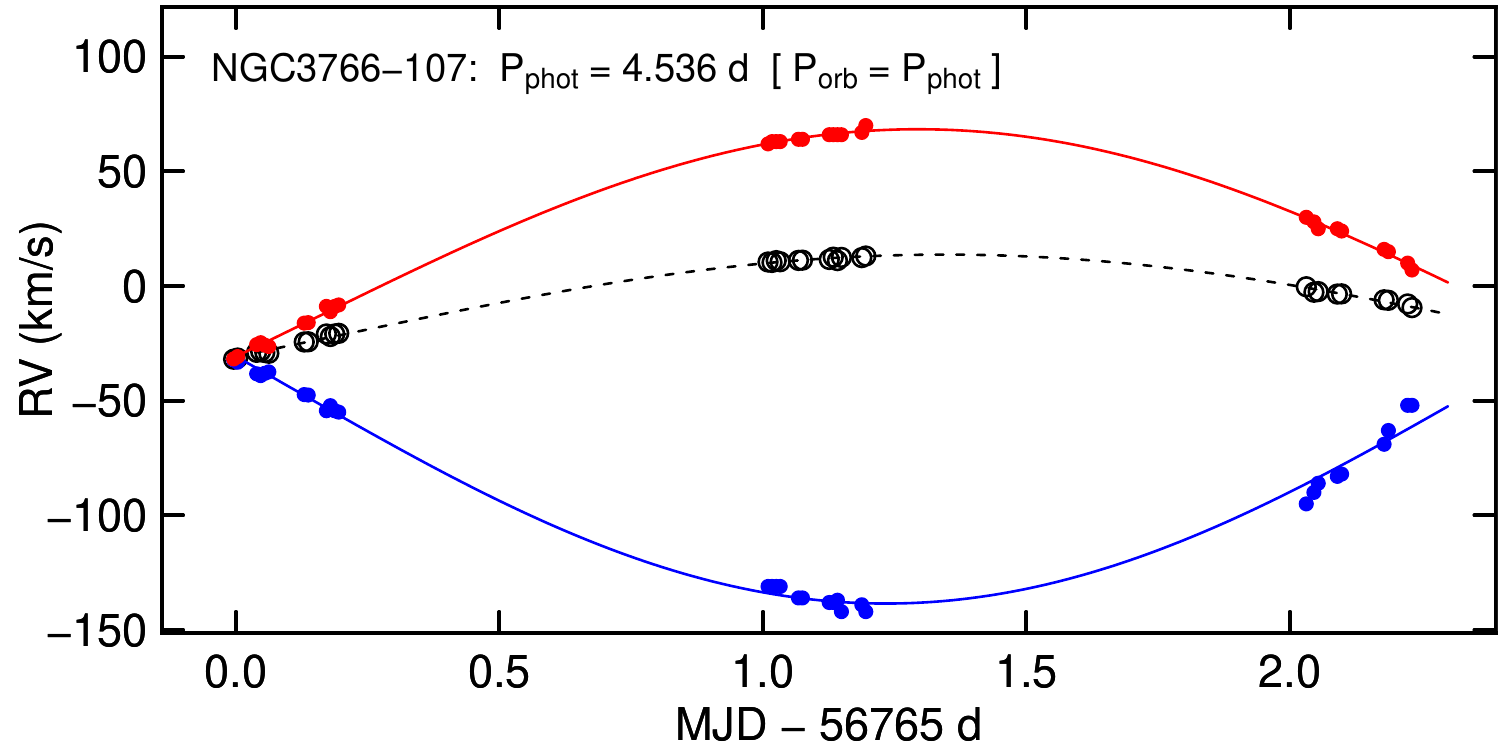}
  \caption{Radial velocity curves of SB2 star 107 in NGC~3766.
           The photometric period is written at the top of the panel next to the star ID and, in parenthesis, the orbital period.
           The RVs of the two components are shown, respectively, red and blue, while the black open circles represent the RVs obtained from a fit of Lorentzian functions to the sides of the H$_\gamma$ and H$_\delta$ lines in the spectra.
           The lines are sine fits to the data using the photometric period.
  }
\label{Fig:RvCurvesSB2}
\end{figure}

Radial velocities (RVs) are computed from the wavelength displacement with respect to laboratory positions of the H$_\gamma$ and H$_\delta$ lines in the LR2 spectra.
The locations of the two lines are computed by simultaneously fitting the sides of the two lines with Lorentzian functions, the sides being defined by the wavelength ranges where spectral fluxes are between 5\% and 80\% of line depth above the minimum flux of each hydrogen line, based on a smoothed approximation of the spectrum.
This method is applicable when the signature of only one star is visible in the spectrum, i.e. for single stars and SB1 binary systems.
When two stellar components are detected in a spectrum (SB2 stars), the H line displacements derived in this way represent a weighted mean of the RVs of the two individual binary system components.
This is shown in Fig.~\ref{Fig:RvCurvesSB2} for SB2 star 107, in which the weighted mean RVs are plotted with dashed line-connected open circles.
The RVs of each stellar component are then computed with a more complex procedure by fitting each individual LR2 spectrum to a grid of synthetic spectra, using a least squares method \citep{SemaanHubertZorec_etal13,FrematNeinerHubert_etal06}.
The resulting stellar component RVs are shown by solid line-connected filled circles in Fig.~\ref{Fig:RvCurvesSB2}.
Typical RV uncertainties are of 6-7~km/s for all our stars.

\begin{figure}
  \centering
  \includegraphics[width=1\columnwidth]{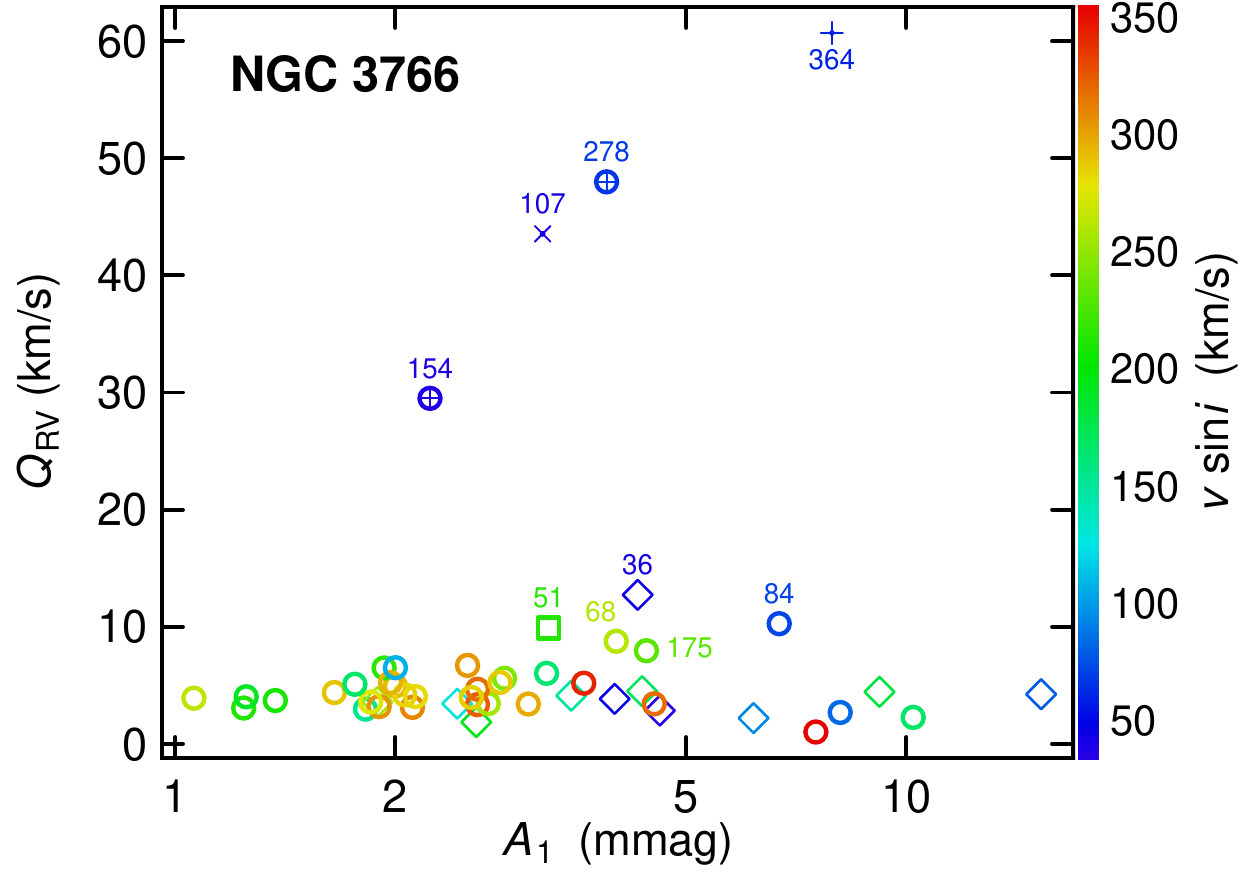}
  \caption{Radial velocity variability amplitude versus photometric variability amplitude of periodic variables in NGC~3766 having VLT/GIRAFFE spectra taken in this program.
           The RV variability amplitude $Q_\mathrm{RV}$ is computed using the 90\%-10\% inter-percentile range.
           Indicator symbols and colors have the same meanings as in Fig.~\ref{Fig:periodMagnitude}.
           Stars showing an RV variability amplitude larger than 7~km/s are labeled with their star IDs next to the indicators.
  }
\label{Fig:RvVarAmplitude}
\end{figure}

\begin{figure}
  \centering
  \includegraphics[width=1\columnwidth]{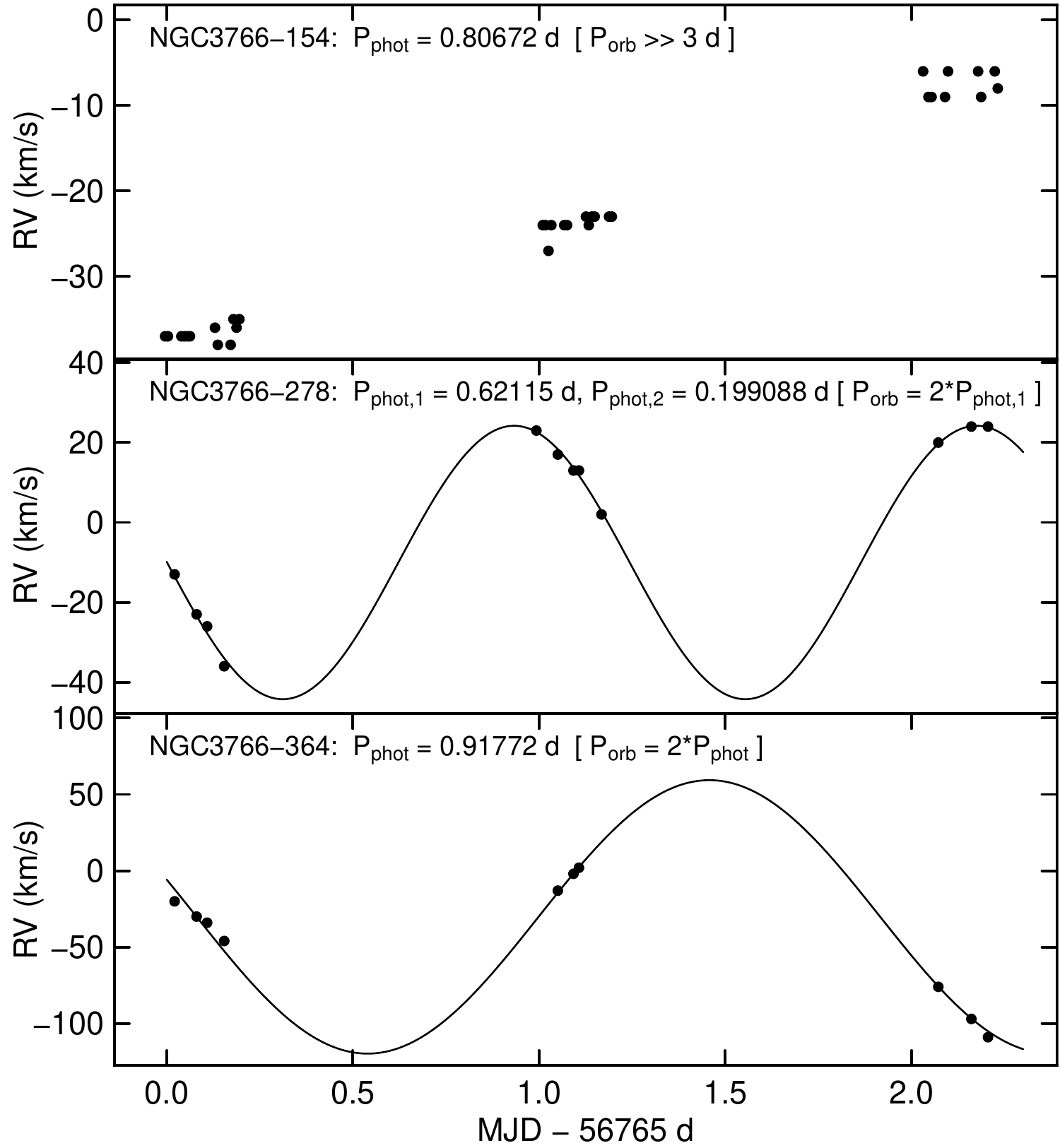}
  \caption{RV curves of three photometric periodic variable stars in NGC~3766 that are confirmed binary systems from their radial velocities.
           The star IDs are written at the top of each panel, together with the photometric periods and, in parenthesis, the orbital period.
           The lines are sine fits to the data using the photometric period.
  }
\label{Fig:RvCurvesSB1}
\end{figure}

The presence of two stellar components is observed in the spectra of only one of our periodic variable targets, Star 107.
The analysis of the RV time series shows a clear variability on a time scale that is compatible with the 4.536~d photometric period (Fig.~\ref{Fig:RvCurvesSB2}).
The star is thus not an SPB star, as initially classified in Paper~I, but an ellipsoidal variable with photometric variability induced by the orbital motion of two tidally distorted stars.

The identification of binaries that are not of SB2 type is more difficult.
We use the RV variation amplitude as a criterion for this purpose.
Amplitudes up to several hundreds of km/s are expected from binary systems with orbital periods in the range 0.1 to 1~days.
The amplitudes expected for SPB and $\beta$ Cep stars fall below $\sim$7~km/s and $\sim$20~km/s, respectively \citep{DeCat02}.
Therefore, any RV variation amplitude greater than about 20~km/s points to the presence of a binary system.

We quantify variability amplitudes by the 90\% - 10\% inter-percentiles of RV curves, which we denote $Q_\mathrm{RV}$.
In addition to the SB2 star 107, three other stars have $Q_\mathrm{RV} > 20$~km/s (Fig.~\ref{Fig:RvVarAmplitude}), all three having been classified in 2013 as new variability class candidates from their photometric properties.
The first of the three stars, star 154, has a photometric period (0.80672 days) that is not related to its orbital period (which is much longer than the two-night baseline of our spectroscopic observations, see Fig.~\ref{Fig:RvCurvesSB1}).
The second star, star 278, has two independent photometric periods (0.199088~d and 0.62115~d), the longest of which is due to orbital motion.
These two stars are thus binary systems with their bright component being a new variability class candidate.
Finally, the RV variability of the third star, mono-periodic star 364, is compatible with its 0.91772~d photometric period (Fig.~\ref{Fig:RvCurvesSB1}).
The star therefore does  not belong to the new variability class, as initially classified in Paper~I, but is an ellipsoidal variable.

In summary, four periodic variables fainter than the bulk of SPB candidates in NGC 3766, Stars 107, 154, 278, and 364, turn out to be binary systems.
They all have photometric periods greater than 0.55~d.
Interestingly, two of them, stars 154 and 278, harbor intrinsically variable stars of the new variability class, making them valuable targets for additional observations.
The periods of their intrinsic variability are 0.80672~d and 0.199088~d, respectively.

\section{Stellar rotation rates and the period-luminosity relation}
\label{Appendix:rotationRates}

\begin{figure}
  \centering
  \includegraphics[width=1\columnwidth]{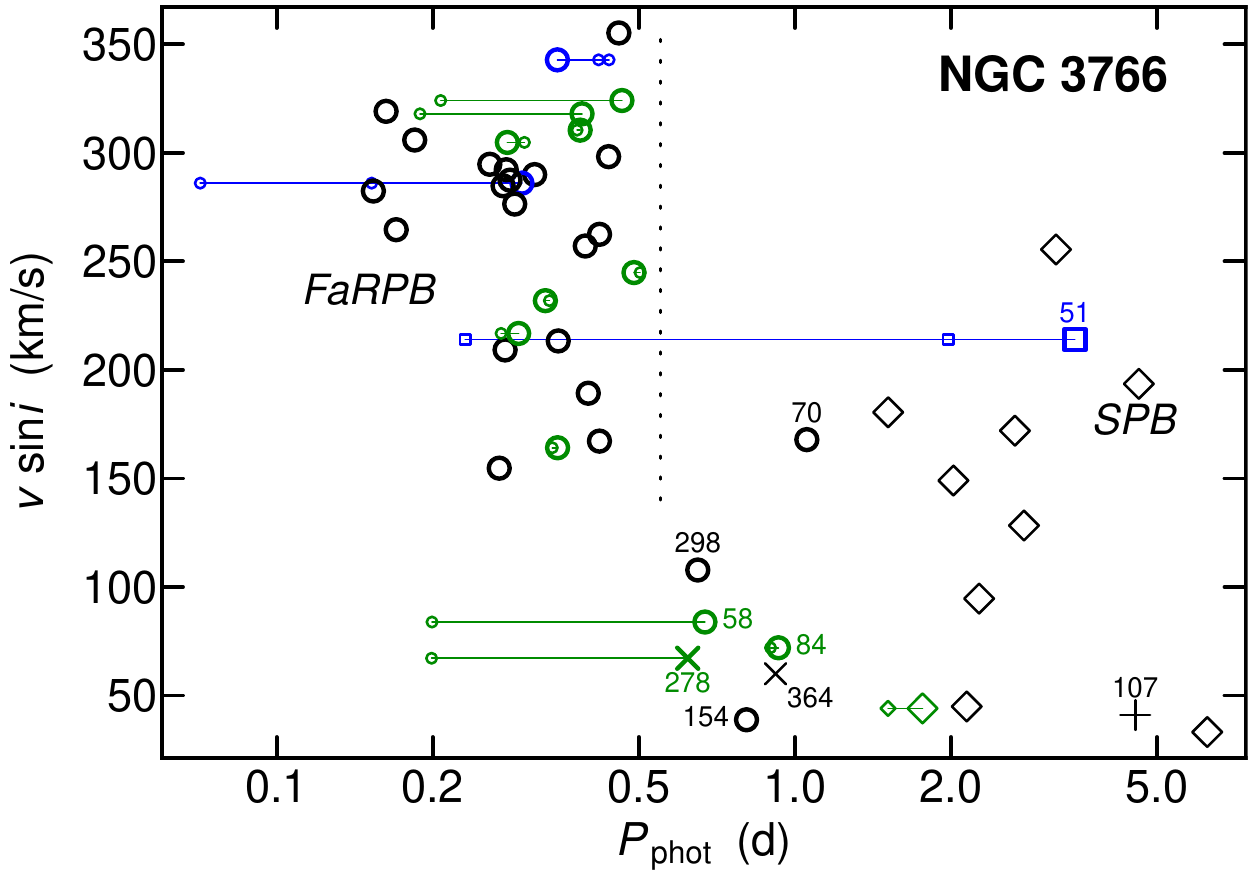}
  \caption{Projected rotational velocity versus photometric periods of SPB (diamonds) and new variability class (circles) candidates in NGC~3766.
           Black, green, and blue inidcators identify mono-, bi-, and tri-periodic stars, respectively.
           Solid horizontal lines link periods within each multiperiodic star, with the second and third periods drawn with smaller indicators.
           Stars labeled in Fig.~\ref{Fig:periodMagnitude} are also reported in this figure.
  }
\label{Fig:periodVSini}
\end{figure}

\begin{figure}
  \centering
  \includegraphics[width=1\columnwidth]{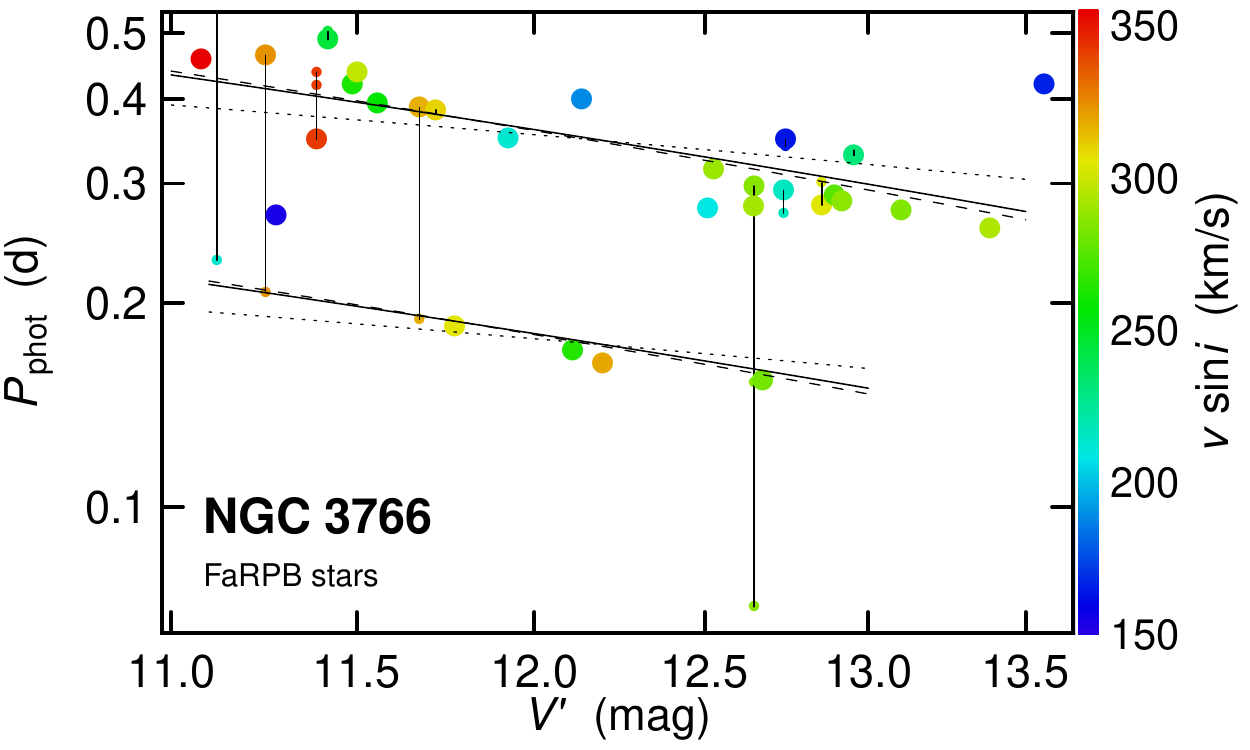}
  \caption{Same as Fig.~\ref{Fig:PLrelation}, but displaying all FaRPB stars.
           The solid lines are fits to all displayed data after multiplication of the second sequence by a factor of two.
           The dashed and dotted lines are identical to the respective lines in Fig.~\ref{Fig:PLrelation}.
           }
\label{Fig:PLrelationAllFaRPBstars}
\end{figure}

\begin{table*}
\centering
\caption{Data of FaRPB stars in NGC~3766.
         $V'$ magnitude uncertainties range from $0.9 \times 10^{-4}$~mag for the brightest stars in the table to $1.2 \times 10^{-4}$~mag for the faintest ones, and typical $v \sin i$ uncertainties are of 10\%.
        }
\begin{tabular}{l c l l l l l}
\hline
StarId &  $V'$   & $P_1$      &  $P_2$      &  $P_3$       & $v\sin i$  \\
       &  (mag)  & (day)      &  (day)      &  (day)       & (km/s)     \\
\hline
 50    & 11.079  &  0.457454(8)&  --         &  --          &  355       \\
 51    & 11.121  &  3.4692(2)  &  1.9777(3)  &  0.23111(0)  &  214       \\
 59    & 11.252  &  0.46378(3) &  0.207271(9)&  --          &  324       \\
 60    & 11.280  &  0.268853(4)&  --         &  --          &  155       \\
 62    & 11.390  &  0.348156(1)&  0.41805(8) &  0.43796(7)  &  343       \\
 66    & 11.421  &  0.48931(2) &  0.50270(8) &  --          &  245       \\
 68    & 11.488  &  0.41983(4) &  --         &  --          &  262       \\
 69    & 11.501  &  0.43710(9) &  --         &  --          &  298       \\
 71    & 11.557  &  0.39406(5) &  --         &  --          &  257       \\
 78    & 11.674  &  0.38845(8) &  0.189088(1)&  --          &  318       \\
 79    & 11.720  &  0.38522(7) &  0.37997(0) &  --          &  310       \\
 83    & 11.773  &  0.184685(8)&  --         &  --          &  306       \\
 94    & 11.925  &  0.34949(7) &  --         &  --          &  213       \\
105    & 12.111  &  0.170123(7)&  --         &  --          &  265       \\
106    & 12.137  &  0.39929(4) &  --         &  --          &  189       \\
112    & 12.198  &  0.162641(8)&  --         &  --          &  319       \\
135    & 12.509  &  0.27596(4) &  --         &  --          &  209       \\
136    & 12.527  &  0.31470(2) &  --         &  --          &  290       \\
142    & 12.649  &  0.29739(5) & 0.0712331(9)&  0.152541(7) &  286       \\
144    & 12.648  &  0.27736(7) &  --         &  --          &  292       \\
145    & 12.675  &  0.153684(4)&  --         &  --          &  282       \\
147    & 12.745  &  0.34838(4) &  0.34033(4) &  --          &  164       \\
149    & 12.739  &  0.29308(2) &  0.27112(6) &  --          &  217       \\
161    & 12.856  &  0.278763(2)&  0.30067(7) &  --          &  305       \\
167    & 12.895  &  0.28800(3) &  --         &  --          &  276       \\
170    & 12.918  &  0.282046(6)&  --         &  --          &  287       \\
175    & 12.955  &  0.330029(6)&  0.33597(0) &  --          &  232       \\
194    & 13.103  &  0.27351(3) &  --         &  --          &  285       \\
236    & 13.385  &  0.25785(1) &  --         &  --          &  295       \\
259    & 13.560  &  0.41966(4) &  --         &  --          &  167       \\
\hline
\end{tabular}
\label{Tab:data}
\end{table*}

Projected rotational velocities are determined from spectral line widening by fitting LR2 spectra to the grid of synthetic spectra.
Typical $v \sin(i)$ uncertainties are of 10\% for fast-rotating stars.

Two groups of stars emerge from a comparison of the projected rotational velocities of SPB and new variability class candidates with their photometric periods (Fig.~\ref{Fig:periodVSini}).
A first group comprises SPB stars and all new variability class candidates with periods larger than 0.55 d.
This first group occupies a well-defined region in the $v \sin(i)$ versus $P_\mathrm{phot}$ plane, starting with slowly rotating new variability class candidates at photometric periods between 0.55~d and 1~d, and extending to SPB stars at higher photometric periods, with a range of projected rotational velocities increasing with period.

The second group of stars consists of new variability class candidates with photometric periods below 0.55~d.
The majority of them, 86\%, are fast rotating with $v \sin(i)>200$~km/s, (and 69\% have $v \sin(i)>250$~km/s).
They form the new class of FaRPB stars that obey the new \PL\ relation presented in Sect.~\ref{Sect:PLrelation} of the main body of the article.

The remaining 14\% FaRPB stars with $v \sin(i)<200$~km/s follow, in general, the \PL\ relation less well.
This is seen in Fig.~\ref{Fig:PLrelationAllFaRPBstars}, which displays the data of all FaRPB stars.
In particular, two of the three stars with $v \sin(i)<170$~km/s clearly deviate from the relation.
Yet, a linear fit to these data still leads to a slope of $(-0.08 \pm 0.01)/mag$.

The data of all FaRPB stars displayed in Fig.~\ref{Fig:PLrelationAllFaRPBstars} are listed in Table~\ref{Tab:data}.
The photometric data are taken from Paper~I, which we repeat here for convenience.

\section{Stellar mass-radius-luminosity dependencies}
\label{Appendix:MRLrelations}

\begin{figure}
  \centering
  \includegraphics[width=1\columnwidth]{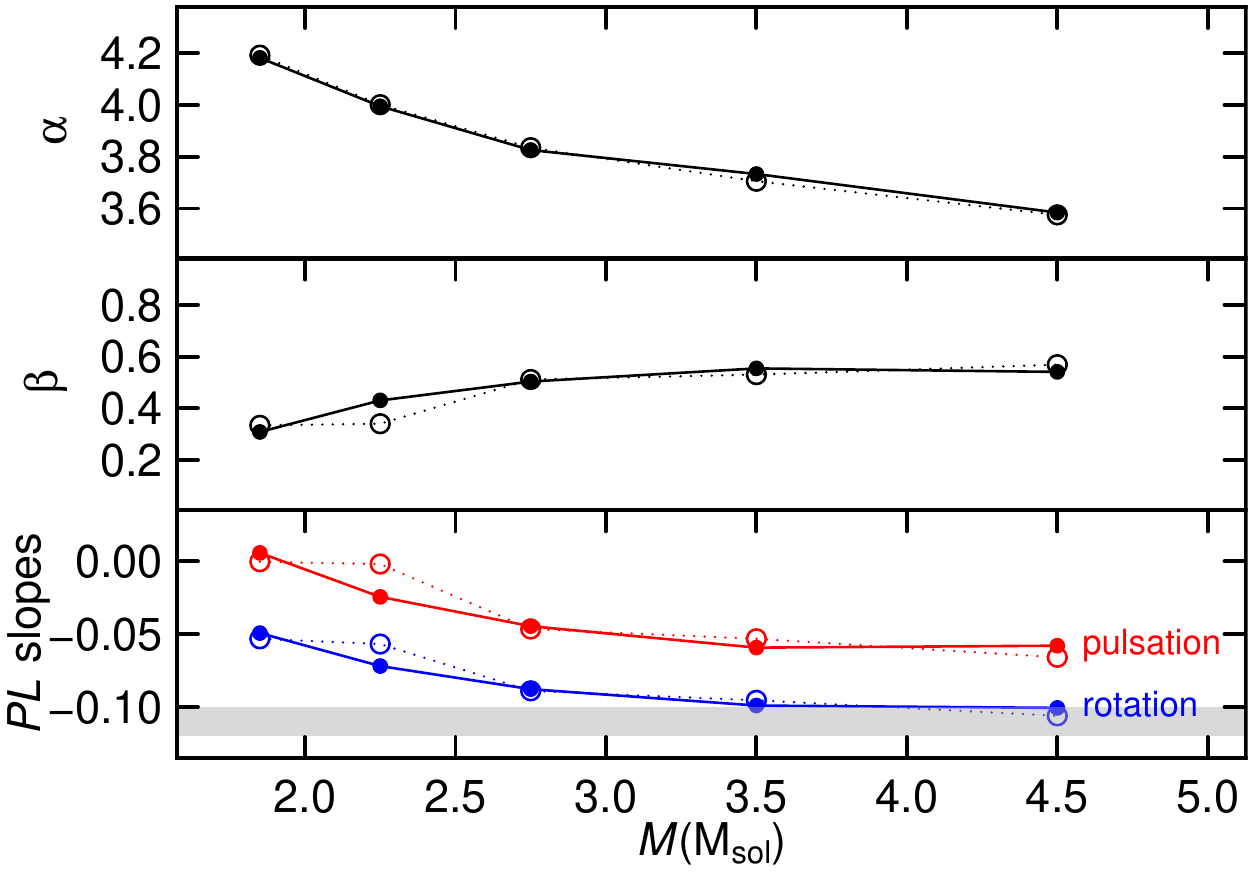}
  \caption{\textbf{Top panel:} $\alpha$ exponent ($L \propto M^\alpha$) of zero-age main sequence stars derived from stellar models of \cite{GeorgyEkstromGranada_etal13} computed without rotation (filled circles connected with dotted lines) or with a near-critical rotation rate corresponding to an angular momentum ratio $\Omega/\Omega_\mathrm{crit}=0.95$ (open circles connected with solid lines).
           \textbf{Middle panel:} Same as top panel, but for the $\beta$ exponent ($R \propto M^\beta$).
           \textbf{Bottom panel:} Slopes of \PL\ relations (in units of 1/mag) using Eqs.~\ref{Eq:Prot} (lower blue lines) and \ref{Eq:Ppuls} (upper red lines) derived from analytical arguments that assume an origin of the \PL\ relation linked to rotation and Cepheid-like pulsation, respectively.
           The horizontal shaded area represents the slope value of $(-0.11 \pm 0.01)/mag$ obtained from observations.
  }
\label{Fig:MLRdependencies}
\end{figure}

\begin{figure}
  \centering
  \includegraphics[width=1\columnwidth]{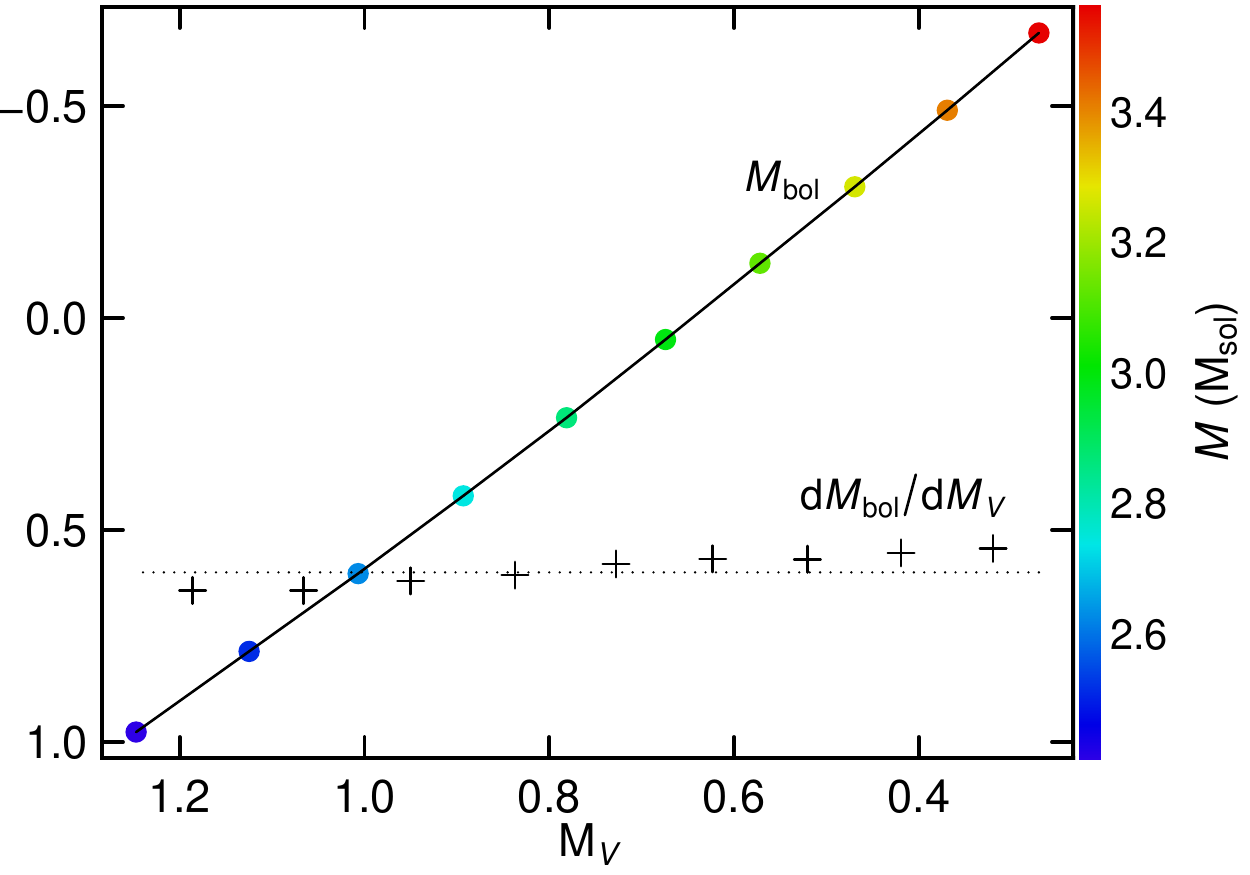}
  \caption{Plot of $M_\mathrm{bol}$ as a function of $M_V$ (filled color circles connected with a line, the colors being related to stellar mass according to the colorscale shown on the right of the figure), from isochrones at $\log(t)=7.4$ ($t$ being the time in yr) published by \citet{GeorgyGranadaEkstrom_etal14}, in the stellar mass range from \mass{2.4} to \mass{3.6}.
           The derivative $dM_\mathrm{bol} / dM_V$ computed numerically between two adjacent data points is plotted with $+$ symbols.
           A dotted line is drawn at the Y-axis value of 0.6, which is the value adopted for this derivative in the analytical considerations presented in the paper.
  }
\label{Fig:MbolMV}
\end{figure}

The $L \propto M^\alpha$ and $R \propto M^\beta$ dependencies of zero-age main sequence stars are derived from solar-metallicity models of rotating stars published by \cite{GeorgyEkstromGranada_etal13}.
The $\alpha$ and $\beta$ exponents are numerically computed from zero-age models using stellar models of masses $M_i = 1.7, 2, 2.5, 3, 4,$ and \mass{5}.
The $\alpha$ exponent is computed as $[\log(L_{i+1})-\log(L_i)] / (M_{i+1}-M_i)$.
The $\beta$ exponent is computed similarly, using stellar radii.
They are shown in Fig.~\ref{Fig:MLRdependencies} for models without rotation and for models with near-critical rotational velocity.
In the typical \mass{2.5-3} mass range of interest to the new variables in NGC~3766, we find $\alpha \simeq 3.8$ and $\beta \simeq 0.5$, with no dependence on stellar rotation rate.

The sensitivity of the \PL\ slopes derived from the analytical relation, given in Eqs.~\ref{Eq:Ppuls} and \ref{Eq:Prot}, is shown in the bottom panel of Fig.~\ref{Fig:MLRdependencies}.
The slopes for both rotational and Cepheid-like pulsation origins are computed as a function of stellar mass using the $\alpha$ and $\beta$ values shown in the top and middle panels, respectively.
A difference of about 0.5 exists at all masses between the slopes derived from rotation and those derived from pulsation, the ones from rotation being close to the value measured from observations given in Eq.~\ref{Eq:PLrelation}, which favors a rotational origin of the relation.

The analytical relation $V \simeq 0.6 M_\mathrm{bol} + \mathrm{cte,}$ used in Sect.~\ref{Sect:origin}, is derived from the isochrones published by \citet{GeorgyGranadaEkstrom_etal14} for solar-metallicity models at $\log(t)=7.4$, $t$ being the time in yr.
The photometric calibrations used by these authors are taken from \citet{WortheyLee11}.
The relation obtained between $M_\mathrm{bol}$ and $M_V$ from those isochrones in the mass range of interest for the FaRPB stars is shown in Fig.~\ref{Fig:MbolMV} (solid line), and the numerically-computed derivative $dM_\mathrm{bol} / dM_V$ from these data points is shown by the $+$ indicators.
The derivative is seen to be equal to about 0.6.

\end{appendix}

\end{document}